\documentclass[aps,twocolumn,titlepage,nofootinbib,longbibliography]{revtex4}
\usepackage{graphicx,physics}% Include figure files
\usepackage{color}
\usepackage{dcolumn}% Align table columns on decimal point
\usepackage{latexsym}
\usepackage{CJKutf8}

\usepackage{hyperref,amssymb}
\usepackage{url}
\usepackage{color,soul}
\usepackage{graphicx}
\usepackage{multirow}
\usepackage{amsmath}
\newcommand{\beq}{\begin{eqnarray}}
\newcommand{\eeq}{\end{eqnarray}}
\usepackage{mathrsfs}
\usepackage{float,soul}
\usepackage[dvipsnames]{xcolor}
\usepackage{mathtools}
\usepackage{slashed}
\usepackage{physics}	% installed packages for typing maths and physics
\usepackage{graphicx}   % need for Fig.ures
\usepackage{epstopdf}
\usepackage{subfigure}  % use for side-by-side Fig.ures
\usepackage{hyperref}   % use for hypertext links, including those to external documents and URLs
\usepackage{bbold}
\usepackage{wasysym}
\usepackage{feynmp}
\usepackage{hyperref}
\hypersetup{colorlinks,
}

\usepackage{xcolor}
\newcommand{\bluedisk}{\textcolor{blue}{\large$\bullet$}}
\newcommand{\magentadisk}{\textcolor{magenta}{\large$\bullet$}}

\begin{document}

\title{\large Rare Events Govern Defect Formation under Weak Symmetry Breaking}
\author{Jiang Liu$^{1,2,3}$}
\author{Peng Yang$^{4,5,1,2}$}
\email{pengyang23@sjtu.edu.cn}
\author{Matteo Baggioli$^{1,2}$}
\email{b.matteo@sjtu.edu.cn}
\address{$^1$Wilczek Quantum Center, School of Physics and Astronomy, Shanghai Jiao Tong University, Shanghai 200240, China}
\address{$^2$Shanghai Research Center for Quantum Sciences, Shanghai 201315, China}
\address{$^3$Tsung-Dao Lee Institute, Shanghai Jiao Tong University, Shanghai 201210, China}
\address{$^4$Department of Physics, The Chinese University of Hong Kong, Shatin, New Territories, Hong Kong 999077, China}
\address{$^5$The State Key Laboratory of Quantum Information Technologies and Materials, The Chinese University of Hong Kong, Shatin, New Territories, Hong Kong 999077, China}

\begin{abstract}
Crossing a continuous phase transition out of equilibrium typically generates topological defects whose density obeys a universal power-law scaling predicted by the Kibble–Zurek mechanism. Recent numerical studies have revealed systematic deviations from this scaling in the presence of weak explicit symmetry breaking, manifested as an additional exponential suppression of defect formation. However, the origin of this correction and a general theoretical framework to describe it have remained elusive. Here, using large-deviation theory, we show that defect formation under weak symmetry breaking is controlled by rare fluctuations that drive local regions into the disfavored symmetry-broken state. Within mean-field theory, this mechanism yields a closed-form expression for the defect density in arbitrary dimensions, valid in the weak-field and weak-noise limits. These theoretical predictions are verified through direct simulations of stochastic Ginzburg–Landau models in one and two spatial dimensions.
\end{abstract}

\maketitle

\section{Introduction}
The formation of topological defects across continuous phase transitions is one of the most striking manifestations of universality far from equilibrium \cite{doi:10.1142/S0217751X1430018X}. It is a paradigmatic example of nonequilibrium critical behavior, in which critical slowing down ties the dynamics to the equilibrium universality class \cite{HohenbergHalperin1977,Tauber2014}. 

When a system is quenched from the symmetric phase into a symmetry-broken state, causally disconnected regions independently select different broken-symmetry vacua, leading to the formation of topological defects. As predicted by the Kibble--Zurek (KZ) mechanism \cite{Kibble_1976,KIBBLE1980183,Zurek1985,Zurek:1993ek}, the density of defects produced during the transition follows a universal power-law dependence on the quench time $\tau_Q$. For sufficiently slow quenches (see \cite{PhysRevLett.130.060402} for the fast quench regime) and defects of zero dimensionality (pointlike), the number of defects scales as
\begin{equation}
n_{\mathrm{KZ}} \propto \tau_Q^{-\frac{d \nu}{1+z\nu}},\label{eq1}
\end{equation}
where $d$ is the spatial dimensionality of the system, and $\nu$ and $z$ are the equilibrium correlation-length and dynamical critical exponents that define the universality class of the transition.

This prediction has been verified in a broad range of experimental platforms, including liquid crystals \cite{Chuang1991,Bowick1994}, superfluid helium \cite{Ruutu1996,Bauerle1996}, colloids \cite{Deutschlander2015}, trapped ions \cite{Ulm2013,Pyka2013}, ultracold atomic gases \cite{Navon2015}, and Rydberg arrays \cite{Keesling2019}; see \cite{doi:10.1142/S0217751X1430018X} for a review. More recently, it has been shown that universality extends beyond the mean defect density: the full counting statistics of defects is also universal, with all cumulants inheriting the KZ scaling \cite{delCampo2018,GomezRuiz2020,King2022}, a result naturally framed within large deviation theory \cite{Touchette2009}. The Kibble--Zurek mechanism is therefore widely regarded as one of the cornerstones of nonequilibrium physics.

More recently, considerable attention has focused on the fate of the Kibble--Zurek mechanism in systems that weakly depart from the paradigm of continuous phase transitions, including weakly first-order transitions \cite{PhysRevLett.132.241601} and transitions subject to weak explicit symmetry breaking \cite{clvs-yk7v}. In both cases, systematic deviations from the standard KZ scaling in Eq.~\eqref{eq1} have been observed. While the former has been successfully explained by combining Kibble--Zurek physics with classical nucleation theory \cite{Langer1969,Oxtoby1992,PhysRevLett.132.241601}, the physical origin of the latter remains poorly understood. In this work, we address this problem and develop a theoretical framework for defect formation under weak explicit symmetry breaking, focusing on the mean defect density.

\begin{figure*}
    \centering
    \includegraphics[width=0.78\linewidth]{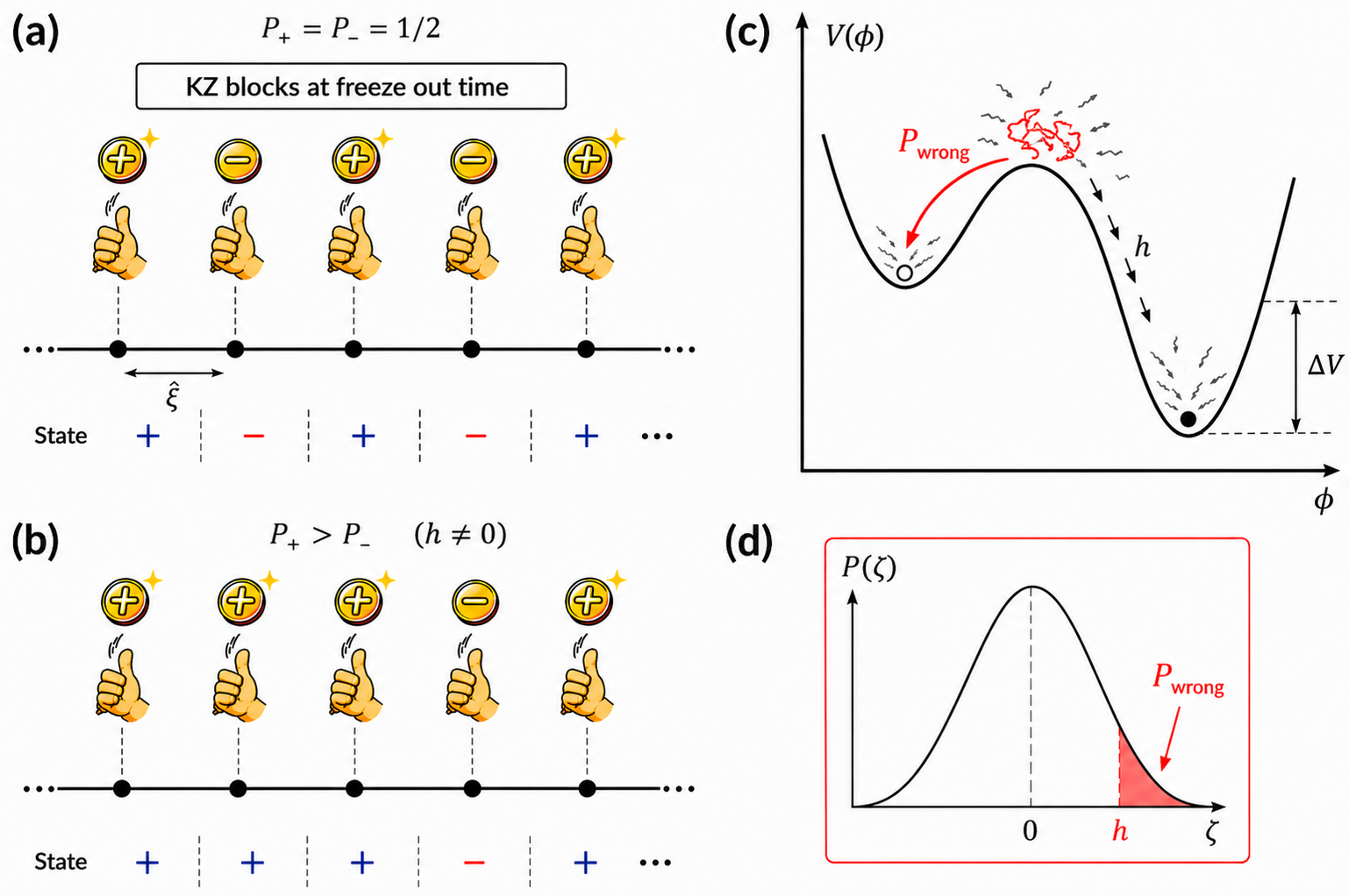}
    \caption{\textbf{Weak Symmetry Breaking as a Biased Coin Toss.} Schematic illustration for the simple case of a $\mathbb{Z}_2$-symmetric system. \textbf{(a)} In the absence of explicit symmetry breaking ($h=0$), independent Kibble–Zurek domains of size $\hat{\xi}$ choose between the two degenerate symmetry-broken states with equal probability, $P_+=P_-=1/2$. \textbf{(b)} A weak external field $h$ biases the local symmetry-breaking choice, leading to unequal probabilities $P_+>P_-$. \textbf{(c)} Defect formation requires a rare fluctuation, occurring with probability $P_{\mathrm{wrong}}$, that drives a Kibble–Zurek domain into the disfavored metastable minimum against the bias imposed by the external field. \textbf{(d)} $P_{\mathrm{wrong}}$ corresponds to the probability of rare fluctuations in the noise distribution that are sufficiently strong to overcome the bias introduced by the external field $h$.}
    \label{fig:1}
\end{figure*}

\section{Physical picture}
In the original KZ framework, a system driven across a continuous phase transition eventually falls out of equilibrium due to critical slowing down. At the freeze-out time $\hat{t}$, the relaxation time becomes comparable to the timescale of the external drive, and the dynamics can no longer follow the quench adiabatically. As a result, the state of the system effectively freezes. The system then breaks up into $N_{\mathrm{blocks}}$ independent correlated regions, or Kibble–Zurek domains, whose characteristic size is set by the correlation length at freeze-out, $\hat{\xi}$ (Fig.~\ref{fig:1}(a)). Since each domain chooses its symmetry-broken state independently and at random, defects form at the interfaces between neighboring domains that make incompatible choices. The resulting defect density is therefore proportional to the number of domains, $
n_{\mathrm{defects}} \propto P_{\mathrm{mismatch}}\,N_{\mathrm{blocks}}$, where $P_{\mathrm{mismatch}}$ is the probability that two neighboring domains select different broken-symmetry states (see Fig.~\ref{fig:1}(a) for the simple case of a $\mathbb{Z}_2$-symmetric system). Because this probability is purely random and independent of the quench rate, it contributes only as a constant prefactor, yielding 
    $n_{\mathrm{KZ}} \propto N_{\mathrm{blocks}}
\propto \hat{\xi}^{-d}.$ Using the critical scaling relations $\hat{t}\sim\tau_Q^{z\nu/(1+z\nu)}$ and $\hat{\xi}\sim\tau_Q^{\nu/(1+z\nu)}$, one immediately recovers the universal scaling form in Eq.~\eqref{eq1}.

However, when a weak symmetry-breaking field $h$ is introduced, the symmetry-breaking choice of each KZ domain is no longer completely random. Instead, the external field biases the local selection of the order parameter, making the process analogous to a biased coin toss that favors the symmetry-broken state corresponding to the globally stable minimum. In the simple case of a $\mathbb{Z}_2$-symmetric system illustrated in Fig.~\ref{fig:1}(b), each KZ domain selects the favored state with probability $P_+$ and the disfavored state with probability $P_-$, with $P_+>P_-$. As a result, the probability of forming a defect is no longer determined by a quench-rate-independent random mismatch between neighboring domains, but by the probability that a domain is driven into the disfavored state against the bias imposed by the external field.

The physical origin of this process is illustrated in Fig.~\ref{fig:1}(c) for a real scalar field $\phi$, where an external field $h$ explicitly breaks the original $\mathbb{Z}_2$ symmetry, tilting the free-energy landscape and lifting the degeneracy between the two symmetry-broken minima, one stable and the other metastable.
 Consequently, the formation of a defect requires a rare fluctuation that overcomes the bias and drives a KZ domain into the disfavored minimum (Fig.~\ref{fig:1}(d)). Denoting this probability by $P_{\mathrm{wrong}} \ll 1$, the defect density becomes $n_{\mathrm{defects}} \propto P_{\mathrm{wrong}}\,N_{\mathrm{blocks}}$, so that the central problem reduces to determining the probability of these rare events.

\section{Theory}
We start by assuming that the standard KZ picture remains valid, such that each correlated domain of size $\hat{\xi}$ independently chooses one of the symmetry-broken states. The corresponding spacetime region over which a single symmetry-breaking decision is made is therefore set by the freeze-out scales, $\Omega_{\rm KZ}
\sim
\hat t\,\hat\xi^d$. We consider stochastic dynamics driven by Gaussian thermal noise,
$\langle
\zeta(\mathbf x,t)
\zeta(\mathbf x',t')
\rangle =2\eta\theta\,
\delta^{(d)}(\mathbf x-\mathbf x')
\delta(t-t')$, for which the probability of a given noise history is
\begin{equation}
P[\zeta]
\propto
\exp\!\left[
-\frac{1}{4\eta\theta}
\int dt\,d^dx\,\zeta^2
\right].
\end{equation}
Here, $\eta$ is the damping coefficient of the overdamped dynamics, which sets the microscopic relaxation time, and $\theta$ is the noise strength, which plays the role of an effective temperature through the fluctuation--dissipation relation. Since $\eta$ sets the microscopic time scale, the freeze-out scales take the form $\hat{t}\sim\eta\,(\tau_Q/\eta)^{z\nu/(1+z\nu)}$ and $\hat{\xi}\sim(\tau_Q/\eta)^{\nu/(1+z\nu)}$, with lengths measured in units of the microscopic length.

As illustrated in Fig.~\ref{fig:1}(c), a KZ domain typically relaxes toward the energetically favored minimum. A domain can end in the disfavored state only if a coherent thermal fluctuation of magnitude $\zeta\sim h$ persists throughout the entire freeze-out spacetime volume. The probability of such a fluctuation is controlled by the tail of the distribution, illustrated schematically in Fig.~\ref{fig:1}(d), and therefore scales as
\begin{equation}
P_{\rm wrong}
\asymp \exp \left[-S_h\right]
=
\exp\!\left[
-\frac{h^2}{4\eta\theta}\Omega_{\rm KZ}
\right],
\end{equation}
where $\asymp$ means equal at leading exponential order.

The corresponding action cost is
\begin{equation}\label{costt}
S_h
\sim
\frac{h^2}{\eta\theta}\,
\hat t\,\hat\xi^d
=
\frac{h^2}{\theta}
\left(
\frac{\tau_Q}
     {\eta}
\right)^{\frac{\nu(d+z)}{1+z\nu}}.
\end{equation}
Since defects can only arise from such rare domains, the defect density is obtained by multiplying this probability by the usual density of independent KZ domains, $n_{\rm KZ}
\sim
\hat\xi^{-d}$, 
which finally yields
\begin{equation}
n_{\rm defects}^{(d)}
\sim
\left(
\frac{\eta}{\tau_Q}
\right)^{\frac{d\nu}{1+z\nu}}
\exp\!\left[
-A_d
\frac{h^2}{\theta}
\left(
\frac{\tau_Q}{\eta}
\right)^{\frac{\nu(d+z)}{1+z\nu}}
\right],\label{maineq}
\end{equation}
where $A_d$ is an overall constant that depends on microscopic details, hence not universal. This is the main result of this work. As for the prefactor of the standard KZ density, the numerical value of $A_d$ depends on order-one ambiguities in the definition of the freeze-out scales and cannot be fixed by the present approach, while its dependence on the microscopic parameters is determined (see SM).

Equation~\eqref{maineq} is expected to hold under three main assumptions. First, it is derived for defects of zero dimensionality, although its generalization to defects of arbitrary dimensionality is straightforward. Second, the explicit symmetry-breaking field $h$ must be sufficiently weak that the underlying KZ freeze-out scales remain essentially unchanged and meaningful. Finally, defect formation must be controlled by rare fluctuations, requiring both weak thermal noise and a large action cost, such that the probability of selecting the disfavored symmetry-broken state is exponentially small.

Importantly, equation~\eqref{maineq} can be put on firmer ground within large-deviation theory, using the Freidlin--Wentzell (FW) action functional~\cite{Freidlin2012}, which gives the probability of a field history in the weak-noise limit $\theta\to0$. The derivation, presented in the Supplemental Material (SM), follows the same steps as the heuristic argument above. Within each Kibble--Zurek domain, the field is taken to be spatially homogeneous over the freeze-out window $[-\hat t,\hat t\,]$, so that the gradient contribution drops out. Because $h$ is weak, the relevant endpoints, namely the initial stable minimum and the boundary of the disfavored basin at freeze-out, are both of order $h$. The dynamics along the near-critical part of the path can therefore be linearized, and the FW action becomes quadratic. Minimizing it under these boundary conditions gives an action proportional to $h^2$, with a coefficient set by the freeze-out volume $\hat t\,\hat\xi^d/(\eta\theta)$ times an order-one number. This reproduces Eq.~\eqref{costt}. The FW analysis thus confirms the exponential factor in Eq.~\eqref{maineq}, but it does not fix the constant $A_d$ or the prefactor, which depend on the nonlinear late-time dynamics and on fluctuation corrections. For the $U(1)$ model, an analogous computation applied to a single-vortex configuration gives the same scaling (see SM).

\begin{figure*}[ht]
    \centering
    \includegraphics[width=0.98\linewidth]{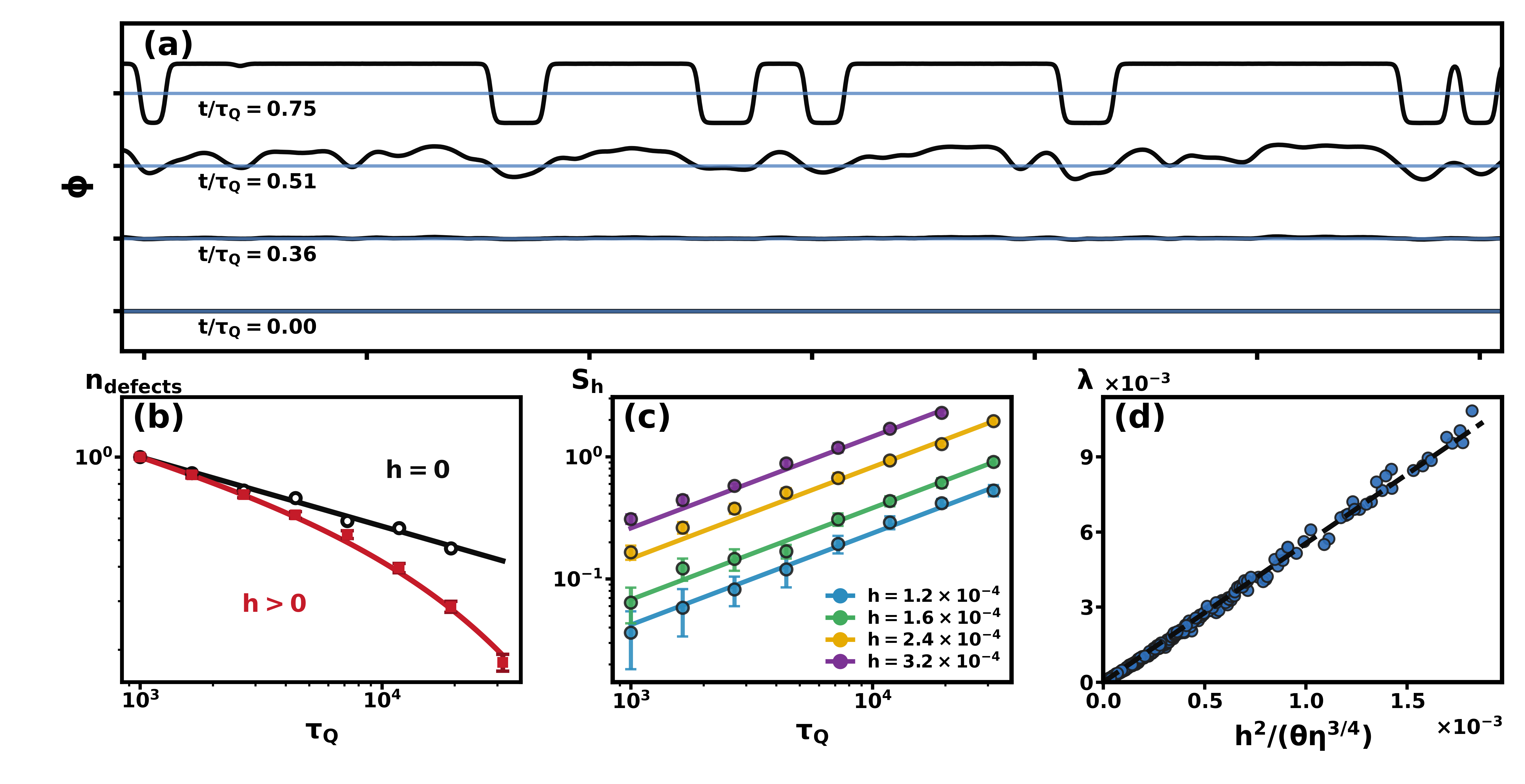}
    \caption{\textbf{Verification of the rare-fluctuation theory in a stochastic 1D Ginzburg--Landau model.}
    \textbf{(a)} Representative snapshots of the order-parameter field during a nonequilibrium quench in the one-dimensional $\mathbb{Z}_2$ model at increasing rescaled times $t/\tau_Q$, with $\tau_Q=5000$, $\eta=100$, $\theta=10^{-6}$, and $h=5\times10^{-5}$. Curves are vertically offset for clarity.
    \textbf{(b)} Normalized defect density at $\eta=100$ and $\theta=10^{-6}$, with $h=0$ (black) and $h=5\times10^{-5}$ (red). Each curve is normalized by its own value at $\tau_Q=1000$. The black curve shows a power-law fit to the unbiased data, while the red curve includes the exponential correction of Eq.~\eqref{maineq}.
    \textbf{(c)} Action cost $S_h$ as a function of $\tau_Q$ at $\eta=100$ and $\theta=10^{-5}$, for the external fields indicated in the legend. Solid lines are fits to the predicted $\tau_Q^{3/4}$ scaling of Eq.~\eqref{maineq}.
    \textbf{(d)} Universal collapse of the parameter $\lambda$ onto the scaling form predicted by Eq.~\eqref{maineq}.
    }
    \label{fig:2}
\end{figure*}

We emphasize that the derivation of Eq.~\eqref{maineq} is fundamentally different from the nucleation-theory corrections arising near weakly first-order transitions \cite{PhysRevLett.132.241601}. In the latter case, rare events are associated with the nucleation of critical droplets that overcome a free-energy barrier. By contrast, in the present scenario defect formation is controlled by rare fluctuations that coherently bias an entire Kibble--Zurek domain toward the disfavored vacuum during freeze-out, without requiring activated crossing of the potential barrier shown in Fig.~\ref{fig:1}(c).

In Ref.~\cite{clvs-yk7v}, the numerically observed exponential correction to the defect density was interpreted in terms of a dynamical correlation length, larger than the equilibrium one. In our interpretation, this scale is naturally identified with the typical distance between the surviving defects, as visible for instance in Fig.~\ref{fig:2}(a). This length scale is larger than the Kibble--Zurek length $\hat{\xi}$ because weak symmetry breaking biases the order parameter toward one of the two sectors, thereby suppressing defect formation and increasing the typical separation between defects in the final non-equilibrium state.

\begin{figure*}[ht]
    \centering
    \includegraphics[width=\linewidth]{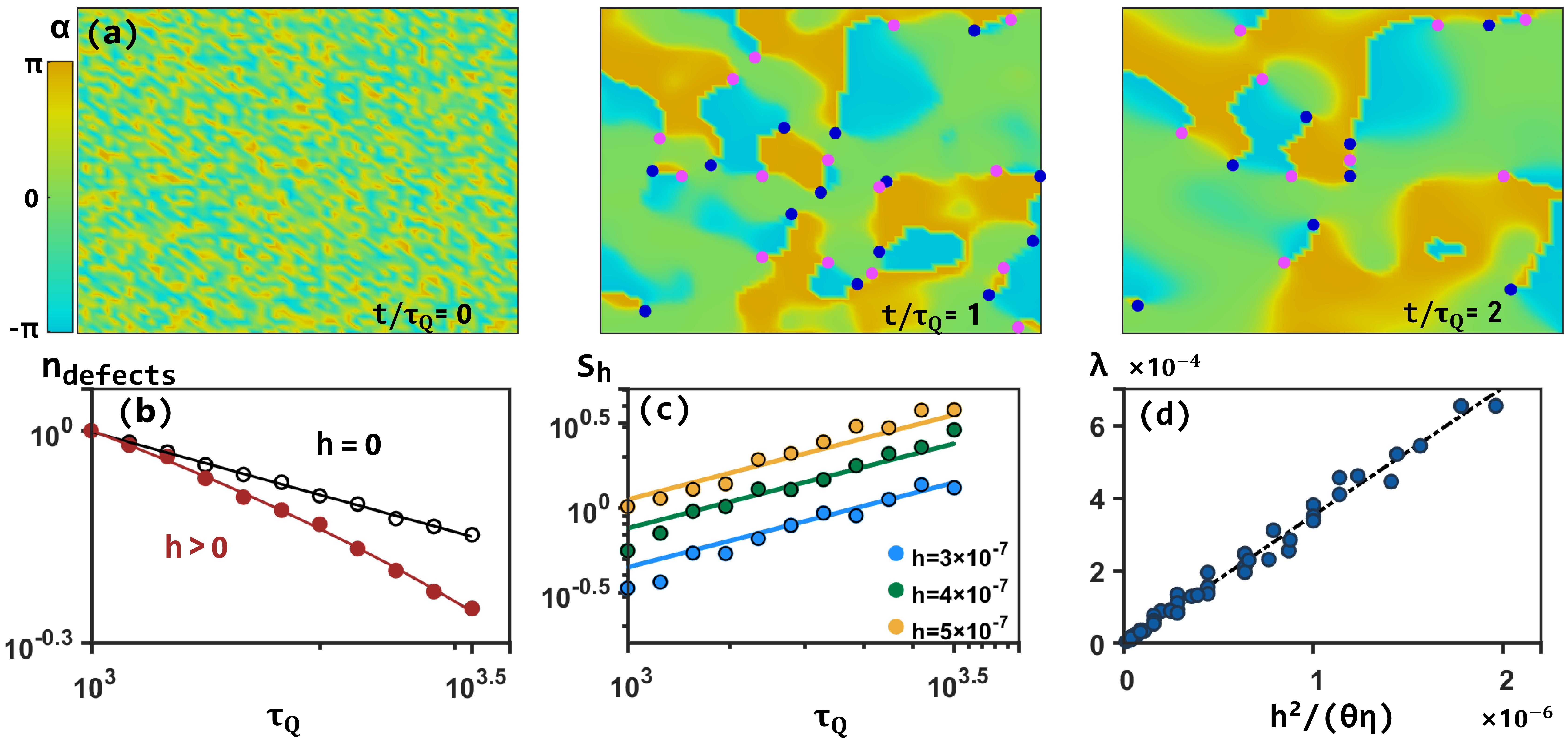}
    \caption{\textbf{Verification of the rare-fluctuation theory in stochastic 2D Ginzburg--Landau model.} \textbf{(a)} Representative
snapshots of the phase field $\alpha(r)$ during nonequilibrium quenches in the two-dimensional $U(1)$ model at increasing rescaled
times $t/\tau_Q$, with $\tau_Q=1000$, $\eta=100$, $\theta=10^{-5}$ and $h=5\times10^{-5}$. Magenta (\magentadisk) and blue (\bluedisk) disks correspond respectively to antivortices and vortices. \textbf{(b)} Normalized defect density at $\eta=100$, $\theta=10^{-5}$ in the 2D Ginzburg--Landau model with ($h=5\times10^{-7}$) and without ($h=0$) an
external symmetry-breaking field $h$. Each curve is normalized by its own value at $\tau_Q = 10^3$. The black curve shows the standard Kibble--Zurek power-law scaling, while the red curve shows $h>0$. \textbf{(c)} Action cost $S_h$ as a function of
$\tau_Q$ at $\eta=100$, $\theta=10^{-5}$ for the 2D model at different values of the external field $h$. Colored solid lines are fits to the predicted $\tau_Q$ scaling, Eq.~\eqref{maineq}. \textbf{(d)} Universal collapse of the parameter $\lambda$ onto the scaling form predicted for the 2D systems, Eq.~\eqref{maineq}.
}
    \label{fig:3}
\end{figure*}

\section{Numerical verification}
To test these predictions, we consider stochastic Ginzburg--Landau dynamics for an order parameter field
$O(\mathbf{x},t)$,
\begin{equation}
\eta\,\partial_t O(\mathbf{x},t)
-\nabla^2 O(\mathbf{x},t)
+\frac{\delta V(O)}{\delta O^\ast(\mathbf{x},t)}
=\zeta(\mathbf{x},t),
\end{equation}
where $\zeta(\mathbf{x},t)$ is the Gaussian thermal noise introduced above, and $\delta V/\delta O^*$ reduces to $\delta V/\delta\phi$ for a real field. We study two representative cases.
The first is a real scalar field $O=\phi$ in one spatial dimension, corresponding to a $\mathbb{Z}_2$
symmetry-breaking transition with domain-wall defects [Fig.~\ref{fig:2}(a)]. The second is a complex scalar field $O=\psi$ in two spatial dimensions, corresponding to a $U(1)$ symmetry-breaking transition with vortex defects [Fig.~\ref{fig:3}(a)].

For the $\mathbb{Z}_2$ model we use
\begin{equation}
V(\phi,t)
=\frac{a(t)}{2}\phi^2
+\frac{b}{4}\phi^4
-h\phi,
\end{equation}
while for the $U(1)$ model we consider
\begin{equation}
V(\psi,t)
=a(t)|\psi|^2
+\frac{b}{2}|\psi|^4
-2h\,\mathrm{Re}\psi.
\end{equation}

In both models, $b>0$ is the quartic coupling, $a(t)=-t/\tau_Q$ is the time-dependent quadratic parameter, and $h$ is a weak symmetry-breaking field. The quadratic parameter is varied linearly in time, with $\tau_Q$ setting the quench timescale. The point $t=0$ corresponds to the
critical point of the unbiased model ($h=0$).
Within mean field, $\nu=1/2$ and $z=2$, so that $\hat{t}=\sqrt{\eta\,\tau_Q}$,
$\hat{\xi}=(\tau_Q/\eta)^{1/4}$, and Eq.~\eqref{maineq} gives $S_h\propto\tau_Q^{(d+2)/4}$.

Details of the two systems, the numerical methods used to solve the non-equilibrium dynamics, and the procedures employed to identify and count topological defects are provided in the SM.

Fig.~\ref{fig:2}(a) shows a representative snapshot of the spatial structure of the order parameter $\phi$ during the nonequilibrium quench in one dimension. Domain-wall defects, corresponding to localized regions where the order parameter switches between the two symmetry-broken states, are clearly visible and originate from rare fluctuations that overcome the bias imposed by the external field.

We numerically compute the defect density generated during nonequilibrium quenches and determine its scaling behavior in the slow-quench regime by averaging over many independent realizations. We observe a clear exponential correction to the standard KZ power-law scaling, consistent with the numerical findings of Ref.~\cite{clvs-yk7v} [Fig.~\ref{fig:2}(b)].

We find that the action cost $S_h$, extracted as $S_h=-\ln(n_h/n_0)$ with $n_0$ the unbiased density at the same $\tau_Q$ and governing the exponential suppression of defects, exhibits the scaling $S_h \propto \tau_Q^{(d+2)/4}$ predicted by Eq.~\eqref{maineq} [Fig.~\ref{fig:2}(c)]. To test the theory quantitatively, we parameterize the defect density as $n_{\rm defects}=n_{\rm KZ}
\exp\!\left[
-\lambda\,\tau_Q^{(d+2)/4}
\right]$, where $\lambda$ is the suppression coefficient. We then extract $\lambda$ over a broad range of symmetry-breaking fields $h$, noise strengths $\theta$, and damping coefficients $\eta$. As predicted by Eq.~\eqref{maineq}, all data collapse onto a single master curve when $\lambda$ is plotted as a function of the scaling variable $h^2/\bigl(\theta\,\eta^{(d+2)/4}\bigr)$, as shown in Fig.~\ref{fig:2}(d).

The same analysis for the two-dimensional $U(1)$ model is summarized in Fig.~\ref{fig:3}. Panel (a) shows snapshots of the phase $\alpha$ at $t/\tau_Q=0,1,2$. Vortices and antivortices form as the phase domains coarsen. As in one dimension, the defect density departs from the Kibble--Zurek power law when $h>0$ [Fig.~\ref{fig:3}(b)]. Since $(d+2)/4=1$ for $d=2$, Eq.~\eqref{maineq} predicts an action cost linear in $\tau_Q$, which is confirmed in Fig.~\ref{fig:3}(c) for several values of $h$. Finally, the suppression coefficient $\lambda$ collapses onto a single line as a function of $h^2/(\theta\eta)$ [Fig.~\ref{fig:3}(d)].

 \section{Outlook}
 The agreement between theory and simulations is excellent in both one and two dimensions, providing strong evidence that weak symmetry breaking transforms defect formation from a Kibble–Zurek process into a rare-event problem. These findings are directly relevant to realistic phase transitions, where perfect symmetry is rarely realized in practice.

In particular, symmetry-violating biases have already been shown experimentally to suppress the Kibble--Zurek mechanism in quantum superfluids~\cite{PhysRevLett.127.115702}, and analogous effects can be engineered in quantum phase transitions using external bias fields~\cite{PhysRevLett.123.130603}. Our theory should therefore apply to these settings and be testable in current experimental platforms.

Finally, while the present analysis is restricted to the mean-field approximation, extending this framework to include beyond-mean-field corrections remains an appealing avenue for future work.

\section*{Acknowledgments}
We thank H.B.~Zeng for useful comments and G.~Volovik for pointing out Ref.~\cite{PhysRevLett.127.115702} to us. JL and MB acknowledge the support of the Shanghai Municipal Science and Technology Major Project (Grant No. 2019SHZDZX01). MB is supported by the Yangyang Development Fund. JL is supported by the T.D. Lee scholarship. PY is partially supported by the National Natural Science Foundation of China (Grant No. 12405021), the Hong Kong Scholar Program (Grant No. XJ2025005), and the China Postdoctoral Science Foundation (Grant No. 2024T170545). % 

\newpage
\onecolumngrid
\appendix 
\clearpage
\renewcommand\thefigure{S\arabic{figure}}    
\setcounter{figure}{0} 
\renewcommand{\theequation}{S\arabic{equation}}
\setcounter{equation}{0}
\renewcommand{\thesubsection}{SM\arabic{subsection}}
\renewcommand{\thesection}{S\arabic{section}}

\makeatletter
\@removefromreset{equation}{section}
\makeatother

\section*{\Large Supplemental Material}

In this Supplemental Material, we provide a formal re-derivation of our main result, Eq.~\eqref{maineq} of the main text, using large-deviation theory and, in particular, the Freidlin--Wentzell action functional within the mean-field approximation. We also provide additional details on the models, simulations, and numerical methods.

\section{Large deviation theory}
\subsection{Freidlin--Wentzell action functional}
In the weak additive-noise limit, fluctuations of a stochastically perturbed dynamical system are governed, at leading exponential order, by Large Deviation Theory (LDT) \cite{SciPostPhysLectNotes.104,Freidlin2012}. In this framework, the probability of observing a prescribed non-equilibrium history is controlled by the Freidlin--Wentzell (FW) action \cite{Freidlin1970}. For a finite-dimensional additive-noise process
\[
\dot X = r(X) + \epsilon \dot W,
\]
the FW estimate for a path $x(t)$ over the time interval $[0,T]$ reads
\[
\mathcal{P}[X(t)\simeq x(t)]
\asymp
\exp\left[
-\frac{1}{2\epsilon^2}
\int_0^T dt\, \left\| \dot x - r(x) \right\|^2
\right],
\]
where the symbol $\asymp$ signifies asymptotic equivalence at leading exponential order.

Thus, the action is the quadratic cost of the noise realization needed to make the system follow the path $x(t)$: deterministic trajectories have $\dot x = r(x)$, while deviations from this drift are exponentially suppressed. For the field theory below, the same idea gives the path probability of a field history $\phi(\mathbf{x},t)$; the only change is that the cost is integrated over both time and space.

For our concrete purpose, we start from the overdamped Langevin equation for a real scalar field $\phi(\mathbf{x},t)$,
\begin{equation}
\eta\partial_t\phi
=
-
\frac{\delta F}{\delta\phi}
+
\zeta(\mathbf x,t),
\label{eq:review-real-langevin}
\end{equation}
with the noise correlation given by
\begin{equation}
\left\langle
\zeta(\mathbf x,t)\zeta(\mathbf x',t')
\right\rangle
=
2\eta\theta\,
\delta^{(d)}(\mathbf x-\mathbf x')\delta(t-t').
\label{eq:review-real-noise}
\end{equation}
By identifying $\zeta(\mathbf x,t)dt=\sqrt{2\eta\theta}\,dW(\mathbf x,t)$ and dividing Eq.~\eqref{eq:review-real-langevin} by $\eta$, we obtain the standard-noise form
\begin{equation}
\partial_t\phi
=
-
\frac{1}{\eta}
\frac{\delta F}{\delta\phi}
+
\epsilon\,\dot W(\mathbf x,t),
\qquad
\epsilon^2=\frac{2\theta}{\eta}.
\label{eq:review-field-sde}
\end{equation}
Here, $\epsilon$ is not an independent physical parameter; rather, it represents the effective small-noise amplitude that arises from rewriting the thermal fluctuations in terms of a unit white noise.

The deterministic drift is
\[
r[\phi]=-\frac{1}{\eta}\frac{\delta F}{\delta\phi},
\]
so the mismatch $\partial_t\phi-r[\phi]$ is the noise history required to generate the path $\phi(\mathbf x,t)$.

The path probability therefore has the asymptotic form
\begin{equation}
\mathcal{P}[\phi]\asymp \exp\left[-S_{\rm FW}[\phi]\right],
\end{equation}
with
\begin{equation}
S_{\rm FW}[\phi]= 
\frac{1}{2\epsilon^2}
\int dt\,d^d x
\left[
\partial_t\phi
+
\frac{1}{\eta}\frac{\delta F}{\delta\phi}
\right]^2=
\frac{1}{4\eta\theta}
\int dt\,d^d x
\left[
\eta\partial_t\phi
+
\frac{\delta F}{\delta\phi}
\right]^2.
\label{eq:review-real-fw}
\end{equation}
Equation~\eqref{eq:review-real-fw} is the central FW result needed below. It states that, in the weak-noise limit $\theta\to0$, the dominant contribution to a rare transition comes from the field history that minimizes this action subject to the relevant boundary or topological constraint. The model-specific noise fields and the defect-producing histories relevant to our problem are discussed in the following subsections. Prefactors and other subleading corrections are not controlled by this leading exponential approximation.

\subsection{\(\mathbb{Z}_2\) model}

We consider a real scalar order parameter $\phi(\mathbf{x},t)$ governed by
overdamped stochastic dynamics with Ginzburg--Landau-type evolution:
\begin{equation}
\eta\partial_t\phi(\mathbf{x},t)-\nabla^2\phi(\mathbf{x},t)
+\partial_\phi V(\phi,t)=\zeta(\mathbf{x},t),
\end{equation}
where $\eta$ is the damping coefficient. The time-dependent potential is
\begin{equation}
V(\phi,t)=\frac{a(t)}{2}\phi^2+\frac{b}{4}\phi^4 -h\phi,
\qquad b>0,\qquad a(t)=-\frac{t}{\tau_Q},
\label{eq:z2-potential}
\end{equation}
where $\tau_Q$ is the quench time and $h>0$ favors the positive phase. The
thermal noise obeys
\begin{equation}
\left\langle\zeta(\mathbf{x},t)\zeta(\mathbf{x}',t')\right\rangle
=2\eta\theta\,\delta^{(d)}(\mathbf{x}-\mathbf{x}')\delta(t-t'),
\end{equation}
with $\theta$ defining the noise-strength scale.

Near the critical point, the linearized homogeneous dynamics obeys
$\eta\dot\phi=-a(t)\phi+h$. The relaxation time and correlation length on
the stable side are
\begin{equation}
\tau_{\rm rel}(t)=\frac{\eta}{|a(t)|}=\frac{\eta\tau_Q}{|t|},
\qquad
\xi(t)=\frac{1}{\sqrt{|a(t)|}}=\sqrt{\frac{\tau_Q}{|t|}}.
\end{equation}
The Kibble--Zurek condition $\tau_{\rm rel}(\hat t)=\hat t$ then yields
\begin{equation}
\hat t=\sqrt{\eta\tau_Q},
\qquad
\hat\xi=\left(\frac{\tau_Q}{\eta}\right)^{1/4}.
\label{eq:z2-kz-scales}
\end{equation}
Thus, this explicit Ginzburg--Landau calculation yields the mean-field exponents of the Gaussian Model-A: $\nu_{\rm MF}=1/2$ and $z_{\rm MF}=2$. We leave the study of beyond-mean-field corrections to future work.

In the absence of a bias ($h=0$), independent correlation domains of volume
$\hat\xi^d$ choose the positive and negative symmetry-broken basins with
equal probability. A finite bias $h>0$ favors the positive basin. Let $p$
denote the rare-event probability that a given domain enters the negative,
energetically disfavored basin. In one dimension, a kink is produced at the
boundary of two adjacent domains that select opposite signs, yielding
$n_{\rm kink}\sim2p(1-p)/\hat\xi$. In the rare-event regime $p\ll1$, this
simplifies to
\begin{equation}
n_{\rm kink}\sim\frac{2p}{\hat\xi}.
\label{eq:z2-kink-density}
\end{equation}

To compute $p$, we employ the Freidlin--Wentzell (FW) action functional.
The path probability scales as $\mathcal{P}[\phi]\asymp
\exp[-S_{\rm FW}[\phi]]$, where
\begin{equation}
S_{\rm FW}[\phi]=\frac{1}{4\eta\theta}\int dt\,d^d x
\left[\eta\partial_t\phi-\nabla^2\phi+b\phi^3+a(t)\phi-h\right]^2.
\label{eq:z2-full-fw}
\end{equation}
During freeze-out ($t\in[-\hat t,\hat t]$), the optimal trajectory is
approximated by the uniform lowest mode within a coherent correlation block
of volume $\hat\xi^d$. Setting $\phi(\mathbf{x},t)\simeq\phi(t)$ eliminates
the gradient term in this block approximation. Neglecting the cubic term
along the early, near-critical segment of the path gives
\begin{equation}
S_{\rm block}[\phi]=\frac{\hat\xi^d}{4\eta\theta}
\int_{-\hat t}^{\hat t}dt\left[\eta\dot\phi-
\frac{t}{\tau_Q}\phi-h\right]^2.
\label{eq:z2-block-fw}
\end{equation}

At the freeze-out endpoints, let
$a_f=|a(\hat t)|=\sqrt{\eta/\tau_Q}$. The initial stable minimum $\phi_i$
at $t=-\hat t$ and the unstable basin boundary $\phi_u$ at $t=\hat t$ are
defined by
\begin{equation}
b\phi_i^3+a_f\phi_i-h=0,
\qquad
b\phi_u^3-a_f\phi_u-h=0.
\end{equation}
For a small external field, namely $\sqrt{b}\,h/a_f^{3/2} \ll 1$, the corresponding solutions at leading order in $h$ are given by
\begin{equation}
\phi_i = \frac{h}{a_f} + O(h^3),
\qquad
\phi_u = -\frac{h}{a_f} + O(h^3).
\label{eq:z2-endpoints}
\end{equation}
We define an incorrect selection event as the system entering the negative basin of attraction, i.e., $\phi(\hat{t}) < \phi_u$. The least-action path reaches this basin boundary in a limiting sense, thereby connecting the initial equilibrium state $\phi(-\hat{t}) = \phi_i$ to the threshold state $\phi(\hat{t}) = \phi_u$.

Introducing the velocity mismatch
$y(t)=\eta\dot\phi-(t/\tau_Q)\phi-h$ and the integrating factor
$\mu(t)=\exp[-t^2/(2\eta\tau_Q)]$, integration over the freeze-out interval
gives
\begin{equation}
\int_{-\hat t}^{\hat t}dt\,\mu(t)y(t)
=\eta\mu(\hat t)\phi_u-\eta\mu(-\hat t)\phi_i
-h\int_{-\hat t}^{\hat t}dt\,\mu(t).
\label{eq:z2-basin-constraint}
\end{equation}
Minimizing $\int y^2dt$ subject to this linear constraint gives
$y_*(t)\propto\mu(t)$ and hence
\begin{equation}
S_h=\frac{\hat\xi^d}{4\eta\theta}
\frac{\left[\eta\mu(\hat t)\phi_u-\eta\mu(-\hat t)\phi_i-
h\int_{-\hat t}^{\hat t}dt\,\mu(t)\right]^2}
{\int_{-\hat t}^{\hat t}dt\,\mu^2(t)}.
\label{eq:z2-sstar}
\end{equation}
Since both $\phi_i$ and $\phi_u$ are proportional to $h$, the quantity in
square brackets in Eq.~\eqref{eq:z2-sstar} is proportional to $h$, and the
squared numerator is therefore proportional to $h^2$. Rescaling time by
$t/\hat t$ shows that the remaining integrals contribute only order-one
numerical constants. Therefore, in $d$ spatial dimensions, the action cost scales as
\begin{equation}
S_h^{(d)}\sim\frac{h^2}{\eta\theta}\hat t\,\hat\xi^d
\sim\frac{h^2}{\theta}\left(\frac{\tau_Q}{\eta}\right)^{\frac{d+2}{4}}.
\end{equation}
Consequently, the probability of selecting the higher-energy basin scales as
\begin{equation}
p_d\asymp\exp\left[-A_d\frac{h^2}{\theta}
\left(\frac{\tau_Q}{\eta}\right)^{\frac{d+2}{4}}\right],
\label{eq:z2-prob-d}
\end{equation}
where $A_d$ is an order-one constant and prefactors, including the unbiased
domain-selection probability, are absorbed into the scaling relation.
Finally, using $n_{\rm rare}^{(d)}\sim\hat\xi^{-d}p_d$, we obtain
\begin{equation}
n_{\rm rare}^{(d)}\sim\left(\frac{\eta}{\tau_Q}\right)^{\frac{d}{4}}
\exp\left[-A_d\frac{h^2}{\theta}
\left(\frac{\tau_Q}{\eta}\right)^{\frac{d+2}{4}}\right].
\label{eq:z2-density-d}
\end{equation}
For $d=1$, this gives
\begin{equation}
n_{\rm kink}\sim\left(\frac{\eta}{\tau_Q}\right)^{1/4}
\exp\left[-A_1\frac{h^2}{\theta}
\left(\frac{\tau_Q}{\eta}\right)^{3/4}\right],
\end{equation}
which is the form used in the main text.

The FW principle controls the leading weak-noise exponential. The uniform,
near-critical block reduction fixes its scaling form up to nonuniversal
corrections from nonlinear late-time dynamics and fluctuation prefactors.

\subsection{\(U(1)\) model in two dimensions}

We now extend this large deviation framework to a complex scalar Ginzburg--Landau model in $d=2$ dimensions, described by the effective free energy functional:
\begin{equation}
F[\psi,h] = \int d^2x \left[ a(t)|\psi|^2 + \frac{b}{2}|\psi|^4 + |\nabla\psi|^2 - 2{\rm Re}(h^*\psi) \right], \qquad b>0.
\end{equation}
The corresponding overdamped stochastic time-dependent Ginzburg--Landau (TDGL) equation is given by:
\begin{equation}
\eta\partial_t\psi = -\left[ a(t)\psi + b|\psi|^2\psi - \nabla^2\psi - h \right] + \zeta(\mathbf{x},t),
\label{eq:u1-tdgl}
\end{equation}
where $\eta$ is the damping coefficient and $\zeta(\mathbf{x},t)$ is a complex Gaussian white noise. Under this convention, the fluctuation-dissipation relation satisfies:
\begin{equation}
\left\langle \zeta^*(\mathbf{x},t)\zeta(\mathbf{x}',t') \right\rangle = 4\eta\theta\,\delta(t-t')\delta^{(2)}(\mathbf{x}-\mathbf{x}'),
\label{eq:u1-noise}
\end{equation}
where $\theta$ defines the thermal noise scale. Writing $\zeta=\zeta_1+i\zeta_2$, we take its real and imaginary
parts to be independent Gaussian white noises, each with covariance $2\eta\theta\,\delta(t-t')
\delta^{(2)}(\mathbf{x}-\mathbf{x}')$, consistently with the real-field noise convention used above. The factor of four in Eq.~\eqref{eq:u1-noise} therefore arises from summing the
two real-component contributions and reflects our chosen normalization of the complex noise. The distance to the continuous phase transition is driven by the linear quench $a(t) = -t/\tau_Q$. In the absence of a bias ($h=0$), the broken-symmetry phase features a continuous ground-state manifold $\psi = \sqrt{-a/b}\,e^{i\alpha}$. A positive real bias field $h > 0$ breaks this $U(1)$ degeneracy, explicitly favoring the $\alpha=0$ direction.

Unlike the discrete $\mathbb{Z}_2$ system, a $U(1)$ topological defect (vortex) cannot be formed by the phase selection of a single independent correlation block. Within a single domain of volume $\hat{\xi}^2$, the phase remains approximately uniform. Instead, a vortex or antivortex is generated when several neighboring blocks spanning a closed loop assemble into a phase configuration characterized by a non-zero integer winding number:
\begin{equation}
W = \frac{1}{2\pi}\oint_{\partial A}\nabla\alpha\cdot d\mathbf{l} = \pm 1.
\label{eq:u1-winding}
\end{equation}
We define the enclosed spatial region bounded by this loop, which scales as $A \sim \hat{\xi}^2$, as a vortex-detecting Kibble--Zurek (KZ) plaquette. Whenever $W = \pm 1$, the phase field possesses a topological singularity, forcing the order parameter amplitude to vanish at the defect core inside the plaquette. 

The mean-field KZ vortex density is established by tiling the system with independent plaquettes of area $\hat{\xi}^2$. In the presence of the bias field, the probability $p_{\rm v}$ for a given plaquette to exhibit a net winding number $W = \pm 1$ is exponentially suppressed. Utilizing the explicit Ginzburg--Landau mean-field exponents $\nu_{\rm MF} = 1/2$ and $z_{\rm MF} = 2$, the standard scaling relations yield the characteristic freeze-out time $\hat{t} \sim \sqrt{\eta \tau_Q}$ and freeze-out correlation length $\hat{\xi} \sim (\tau_Q/\eta)^{1/4}$. The resulting vortex density scales as:
\begin{equation}
n_{\rm v} \sim \hat{\xi}^{-2}p_{\rm v}(h,\theta) \sim \left(\frac{\eta}{\tau_Q}\right)^{1/2} p_{\rm v}(h,\theta).
\label{eq:u1-density}
\end{equation}

To evaluate the plaquette winding probability $p_{\rm v}$, we formulate the Freidlin--Wentzell (FW) action functional for the complex field dynamics:
\begin{equation}
S_{\rm FW}[\psi] = \frac{1}{4\eta\theta} \int dt\,d^2x \left| \eta\partial_t\psi + a(t)\psi + b|\psi|^2\psi - \nabla^2\psi - h \right|^2.
\label{eq:u1-fw}
\end{equation}
Let $\mathcal{C}_{\pm 1}$ denote the set of field histories that terminate in configurations containing a winding number $W = \pm 1$ around the designated plaquette. Path integration yields the asymptotic saddle-point approximation $p_{\rm v}(h,\theta) \asymp \exp[-S_{\rm v}^{\min}]$, where:
\begin{equation}
S_{\rm v}^{\min} = \inf_{\psi\in{\cal C}_{+1}\cup{\cal C}_{-1}} S_{\rm FW}[\psi].
\label{eq:u1-svmin}
\end{equation}

We estimate this constrained minimum via a coarse-grained point-vortex ansatz defined over a domain of radius $R \sim \hat{\xi}$:
\begin{equation}
\psi(r,\vartheta,t) \simeq \psi(t)e^{im\vartheta}, \qquad m=\pm1, \qquad \varepsilon<r<R.
\label{eq:u1-vortex-ansatz}
\end{equation}
Along the near-critical portion of the optimal escape trajectory, the non-linear terms can be neglected. Substituting Eq.~\eqref{eq:u1-vortex-ansatz} into the FW action leads to:
\begin{equation}
S_{\rm FW} \simeq \frac{1}{4\eta\theta} \int_{-\hat{t}}^{\hat{t}}dt \int_\varepsilon^R r\,dr\int_0^{2\pi}d\vartheta \left| e^{im\vartheta} \left[ \eta\dot\psi + a(t)\psi + \frac{ m^2\psi}{r^2} \right] - h \right|^2.
\label{eq:u1-annulus}
\end{equation}
For elementary vortices ($m=\pm 1$), the cross term between the winding phase factor $e^{im\vartheta}$ and the uniform source $h$ vanishes identically upon angular integration. Consequently, the explicit symmetry-breaking contribution isolates cleanly into a distinct energy penalty:
\begin{equation}
S_h \simeq \frac{1}{4\eta\theta} \int_{-\hat{t}}^{\hat{t}}dt\, \pi(R^2-\varepsilon^2)h^2,
\label{eq:u1-source-action}
\end{equation}
which physically represents the energetic cost required to maintain a topologically trapped winding configuration that cannot globally align with the external field. 

Evaluating this integral over the freeze-out temporal window and scaling the boundary as $R \sim \hat{\xi}$ directly provides the action cost. Using the explicit mean-field solutions for $\hat{t}$ and $\hat{\xi}$ to construct the characteristic spacetime volume $\hat{t} \hat{\xi}^2 \sim \tau_Q$, we find that the explicit field contribution scales as:
\begin{equation}
S_h \sim A_2\frac{h^2}{\eta\theta} \hat{t} \hat{\xi}^2 \sim A_2 \frac{h^2 \tau_Q}{\theta \eta},
\label{eq:u1-sh-scaling}
\end{equation}
where $A_2$ is an order-one constant collecting the spatial core cutoffs, stiffness parameters, and profile details. Restoring this to the density relation \eqref{eq:u1-density} yields the final defect density verified in the main text:
\begin{equation}
n_{\rm v} \sim \left(\frac{\eta}{\tau_Q}\right)^{1/2} \exp\left[ - A_2 \frac{h^2 \tau_Q}{\theta \eta} \right].
\label{eq:u1-final-density}
\end{equation}

Remarkably, the $\mathbb{Z}_2$ and $U(1)$ descriptions reveal a unified large-deviation origin: the external field biases the order parameter, and defect formation requires a stochastic fluctuation capable of overcoming this bias throughout the characteristic mean-field freeze-out spacetime volume. While a $\mathbb{Z}_2$ defect arises locally from a single domain choosing the disfavored minimum, a $U(1)$ defect emerges from a collective phase mismatch along a multi-block plaquette. Despite these distinct topological structures, both models collapse onto the same large-deviation structure dictated by the fluctuation-dissipation scale $\theta$, verifying the consistency of the FW framework across different symmetry groups under Ginzburg--Landau dynamics.

\section{Numerical methods}
\subsection{$\mathbb{Z}_2$ model}

The numerical simulations used to test the $\mathbb{Z}_2$ predictions solve the one-dimensional overdamped stochastic differential equation:
\begin{equation}
\eta\partial_t\phi = \partial_x^2\phi - \left( \phi^3 - \frac{t}{\tau_Q}\phi - h \right) + \zeta(x,t),
\label{eq:num-eom}
\end{equation}
subject to periodic boundary conditions. The stochastic term is implemented as Gaussian white noise with a variance matching the fluctuation-dissipation relation:
\begin{equation}
\left\langle \zeta(x,t)\zeta(x',t') \right\rangle = 2\eta\theta\,\delta(x-x')\delta(t-t').
\end{equation}
On a discrete lattice with spatial spacing $dx$ and temporal step $dt$, the stochastic increment is sampled with variance $2\eta\theta\,dt/dx$. The spatial Laplacian is evaluated pseudospectrally via Fast Fourier Transforms (FFTs), and the time evolution is integrated using a second-order stochastic Heun predictor-corrector scheme.

The system size and grid dimensions utilized in the production runs are:
\begin{equation}
L=2048, \qquad N_x=16384, \qquad dx=\frac{L}{N_x},
\end{equation}
with a fixed time step of $dt=0.1$. For each parameter set, the dynamics are evolved over 20 independent noise realizations in parallel. The initial condition consists of a spatially uniform field situated at the positive local minimum of the initial potential, seeded with a small random perturbation to stimulate initial fluctuations.

The theoretical freeze-out time is defined as $\hat{t}=\sqrt{\eta\tau_Q}$. Rather than using $\hat{t}$ directly as the defect-counting threshold, this time scale sets the fundamental Kibble--Zurek length scale and the expected scaling behavior. In the simulations, the field is evolved well past the transition point to ensure that independent domains and sharp kinks are fully developed. The total defect number is subsequently recorded at a late time:
\begin{equation}
t_{\rm count}=2\tau_Q.
\label{eq:num-tcount}
\end{equation}
This delayed counting procedure eliminates ambiguities associated with the frozen near-critical regime and measures the topological defects only after the order parameter has firmly settled into the broken-symmetry basins.

The snapshots presented in Fig.~\ref{fig:2}(a) of the main text display representative field profiles $\phi(x,t)$ from a single quench at several rescaled times $t/\tau_Q$. The curves in that panel are vertically shifted for visual clarity. As the quench progresses deep into the broken-symmetry regime, distinct spatial domains settle into opposite signs of $\phi$; the zeros separating these domains constitute the one-dimensional $\mathbb{Z}_2$ defects.

Defects in the one-dimensional $\mathbb{Z}_2$ simulations are tracked as sign changes of the order parameter between adjacent lattice points. For a final field configuration $\phi_i$, we define the discrete sign variable:
\begin{equation}
\sigma_i = {\rm sign}(\phi_i) = 
\begin{cases}
+1, & \phi_i \ge 0, \\
-1, & \phi_i < 0.
\end{cases}
\end{equation}
The total number of kinks is then computed via:
\begin{equation}
N_{\rm kink} = \sum_{i=1}^{N_x-1} \frac{1-\sigma_i\sigma_{i+1}}{2},
\label{eq:num-defect-count}
\end{equation}
which represents the lattice implementation of the neighboring-block sign-mismatch criterion derived in Eq.~\eqref{eq:z2-kink-density}. The reported defect number for each parameter set is computed as the ensemble average of $N_{\rm kink}$ over all independent stochastic realizations.

\subsection{$U(1)$ model}

To simulate the Kibble--Zurek mechanism in the two-dimensional $U(1)$ system, we consider a periodic box of size $L = 500$ discretized on a square grid of $N_x \times N_y = 500 \times 500$ points ($dx=1$). The stochastic time evolution from the initial state $\psi(\mathbf{x}, t_0)$ to the final state $\psi(\mathbf{x}, t_f)$ is governed by the complex time-dependent Ginzburg--Landau equation given in Eq.~\eqref{eq:u1-tdgl}:
\begin{equation}
\partial_t \psi = \mathcal{L}(a(t), \psi(\mathbf{x},t)) + \frac{1}{\eta}\zeta(\mathbf{x},t),
\end{equation}
where $\mathcal{L}$ represents the deterministic drift terms. The equations are integrated numerically using a fourth-order Runge-Kutta method adapted for stochastic fields, with the system initialized in the ground state of the disordered phase far from the critical point and a fixed time step of $dt=0.1$. The complex Gaussian white noise $\zeta(\mathbf{x},t)$ is generated at each step with variance scaled by $4\eta\theta dt / dx^2$ in accordance with the fluctuation-dissipation relation in Eq.~\eqref{eq:u1-noise}.

The total number of vortices is measured at a late characteristic time scale $t=c\hat{t}$ (where $c$ is an order-one constant). At this stage of the quench, local subdomains of the form $\psi_i = |\psi_i|e^{i\alpha_i}$ are well-formed and the spatially averaged order parameter satisfies $\langle|\psi|\rangle \approx 10\% \langle|\psi(t_f)|\rangle$. The vortices and antivortices are identified on the grid by evaluating the discrete phase winding over elementary four-point plaquettes, calculating the contour integral of the phase gradient to identify singularities where $W = \pm 1$, consistent with the criteria established in Eq.~\eqref{eq:u1-winding}.

\section{Additional test of microscopic-parameter dependence}

We also examine the dependence of the suppression coefficient on two microscopic parameters: the gradient stiffness $\kappa$ and the scale $r_0$ of the quadratic term. We consider
\begin{equation}
\eta\,\partial_t\phi - \kappa\,\partial_x^2\phi + \partial_\phi V(\phi,t) = \zeta(x,t),\label{eee}
\end{equation}
with
\begin{equation}
V(\phi,t) = \frac{a(t)}{2}\phi^2 + \frac{b}{4}\phi^4 - h\phi, \qquad a(t) = -r_0\frac{t}{\tau_Q},\quad b>0,\ r_0>0.
\end{equation}
Since $\kappa$ and $r_0$ can be absorbed by rescaling $x\to x/\sqrt{\kappa}$ and $\tau_Q\to\tau_Q/r_0$, their effect on the suppression coefficient follows from the scaling of Eq.~\eqref{maineq} with $\theta$ and $\tau_Q$. The comparison below is therefore a consistency check of the numerical extraction procedure, which is not invariant under this rescaling because the grid spacing, system size, and counting time are held fixed.

Equation \eqref{eee} gives the mean-field freeze-out scales
\begin{equation}
\hat t=\left(\frac{\eta\tau_Q}{r_0}\right)^{1/2},
\qquad
\hat\xi=\sqrt{\kappa}\left(\frac{\tau_Q}{\eta r_0}\right)^{1/4}.
\end{equation}
The estimate $S_h^{(d)}\sim h^2\hat t\hat\xi^d/(\eta\theta)$ therefore suggests the leading dependence $A_d\sim\widetilde A_d\kappa^{d/2}r_0^{-(d+2)/4}$. In one dimension, writing $S_h=\lambda\tau_Q^{3/4}$ and $K=\lambda/h^2$ yields
\begin{equation}
K\sim\frac{\widetilde A_1}{\theta\eta^{3/4}}\,\kappa^{1/2}r_0^{-3/4}.
\label{eq:sm-microscopic-K}
\end{equation}

\begin{figure}[htbp]
    \centering
    \includegraphics[width=0.7\linewidth]{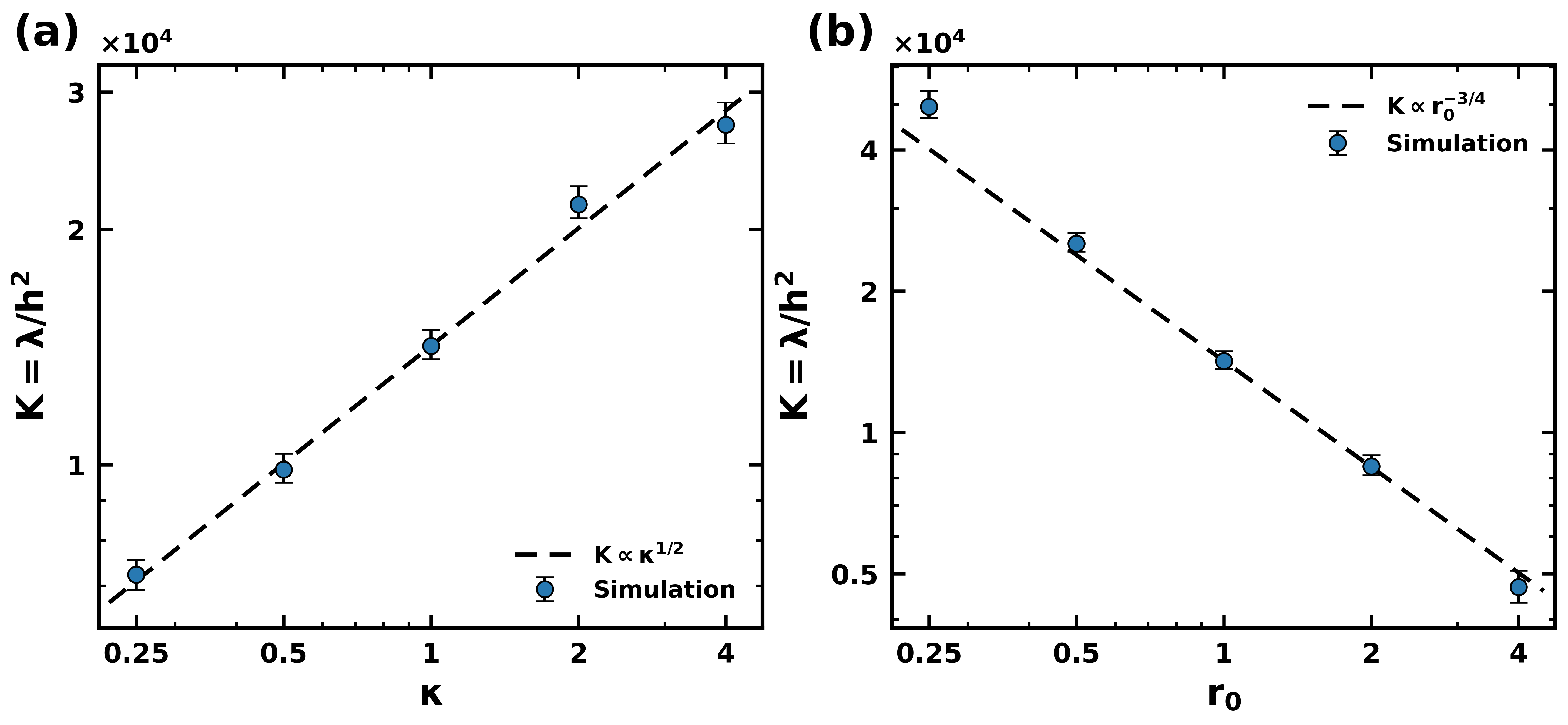}
    \caption{\textbf{Additional test of microscopic-parameter dependence.} Suppression coefficient $K=\lambda/h^2$ in the one-dimensional model. \textbf{(a)} Varying $\kappa$ at fixed $r_0=1$. \textbf{(b)} Varying $r_0$ at fixed $\kappa=1$. Other parameters are fixed at $b=1$, $\eta=100$, and $\theta=10^{-5}$. Dashed curves show $K\propto\kappa^{1/2}$ and $K\propto r_0^{-3/4}$, normalized to the same reference value at $\kappa=r_0=1$. Blue circles denote simulation results; error bars indicate $95\%$ bootstrap confidence intervals.}
    \label{fig:sm-Ad}
\end{figure}

We verify this scaling numerically by varying $\kappa$ and $r_0$ independently, using each parameter set's own unbiased reference to extract the suppression coefficient. The comparison in Fig.~\ref{fig:sm-Ad} confirms that the extracted coefficient follows the expected scaling. Both curves are normalized using the same reference value at $\kappa=r_0=1$, without further amplitude adjustments. The overall numerical factor $\tilde{A}_1$ is not fixed by the scaling argument.

\end{document}